\documentclass[aip,pop,reprint]{revtex4-1}
 \usepackage{amssymb}
 \usepackage{bigints}
 \usepackage{amsmath}
 \usepackage{amsthm}
 \usepackage{empheq}
 \usepackage{framed}
 \usepackage{color}
 \usepackage{enumerate}
 \usepackage{makeidx}
 \usepackage{graphicx}
 \usepackage{ulem}
 \usepackage{ifpdf}
 \usepackage[caption=false]{subfig}
 \usepackage{setspace}

 \usepackage{bm}
 \usepackage{graphics}
 \usepackage{colortbl}
 \usepackage{colordvi}
 \usepackage{verbatim}
 \usepackage{multirow} 
 \usepackage{booktabs} 
 \usepackage{appendix}
 \usepackage{upgreek}
 \usepackage{textgreek}
 \usepackage{mathrsfs}
 
\newcommand{\be}{\begin{equation}}
\newcommand{\ee}{\end{equation}}
\newcommand{\ba}{\[\begin{aligned}}
\newcommand{\ea}{\end{aligned}\]}
\newcommand{\bea}{\begin{eqnarray}}
\newcommand{\eea}{\end{eqnarray}}
\newcommand{\beann}{\begin{eqnarray*}}
\newcommand{\eeann}{\end{eqnarray*}}
\newcommand{\bs}{\begin{split}}
\newcommand{\es}{\end{split}}

\newcommand*\at[2]{\left.#1\right|_{#2}}

\newcommand*{\cB}{\mathcal{B}}

\newcommand*{\cD}{\mathcal{D}}
\newcommand*{\cE}{\mathcal{E}}

\newcommand*{\ep}{\epsilon}

\newcommand*{\B}{\bm{B}}

\newcommand*{\J}{\bm{J}}
\newcommand*{\Jac}{\mathscr{J}}

\newcommand*{\vpl}{v_{||}}

\newcommand*{\dpl}{\nabla_{||}}

\newcommand*{\Jpl}{J_{||}}

\newcommand*{\dt}{\mathrm{d}}
\newcommand*{\dtl}{\mathrm{d}l}

\newcommand*{\dl}{\bm{\nabla}}
\newcommand*{\del}{\partial}

\newcommand*{\xh}{\bm{\hat{x}}}
\newcommand*{\yh}{\bm{\hat{y}}}
\newcommand*{\zh}{\bm{\hat{z}}}

\newcommand*{\zb}{\bar{z}}
\newcommand*{\zbs}{\zb^{(\sigma)}}
\newcommand*{\zbp}{\zb^{(+)}}
\newcommand*{\zbm}{\zb^{(-)}}
\newcommand*{\Bb}{\bar{B}}

\newcommand*{\xcg}{x_{,\gamma}}

\newcommand*{\xcz}{x_{,z}}
\newcommand*{\xcsi}{x_{,\Psi}}
\newcommand*{\ycg}{y_{,\gamma}}
\newcommand*{\ycz}{y_{,z}}
\newcommand*{\ycsi}{y_{,\Psi}}

\newcommand*{\lbr}{\left(}
\newcommand*{\rbr}{\right)}

\newcommand*\smpi[2]{s_{#1}^{#2}}
\newcommand*{\Bmin}{B_{\text{min}}}
\newcommand*{\Bmax}{B_{\text{max}}}
\newcommand*{\wmax}{w_{\text{max}}}

\begin{document}

\title{Radial confinement of deeply trapped particles in a non-symmetric magnetohydrodynamic equlibrium} 

\author{Wrick Sengupta, Harold Weitzner}

\affiliation{Courant Institute of Mathematical Sciences, New York University, New York, New York 10012, USA}

\date{\today}

\begin{abstract}
  Quasisymmetry and omnigeneity of an equilibrium magnetic field are two distinct properties proposed to ensure radial localization of collisionless trapped particles in any stellarator. These constraints are incompletely explored, but have stringent restrictions on a magnetic geometry.  This work employs an analytic approach to understand the implications of the constraints. 
 The particles move in an intrinsically three dimensional equilibrium whose representation is given by earlier work of Weitzner and its extension here. For deeply trapped particles a local equilibrium expansion around a minimum of the magnetic field strength along a magnetic line suffices. This analytical non-symmetric equilibrium solution, enables explicit representation of the constraints.  The results show that it is far easier to satisfy the omnigeneity condition than the quasisymmetry requirement. Correspondingly, there exists a large class of equilibrium close to quasisymmetry that remain omnigeneous while allowing inclusion of error fields, which may destroy quasisymmetry. 
\end{abstract}


\maketitle 

\section{Introduction}
Stellarators offer the attractive paradigm of intrinsic steady state operation. Unlike tokamaks, they neither rely on toroidal currents to provide the rotation transform nor do they appear to suffer from disruptions. Non-axisymmetry and hence full three-dimensional nature of magnetic fields then follows. While the three-dimensional nature of the magnetic geometry allows a wide range of design choices for stellarators, it also has a major disadvantage: in the absence of symmetry there is no guarantee that trapped particles will remain near a given flux surface and thus be confined. To avoid such radial drifts of particles, properties of the magnetic field such as quasisymmetry \cite{nuhrenberg1988quasi,boozer1995quasi} and omnigeneity \cite{hall1975three,caryshashrinaPRL1997helical} have been proposed. A stellarator with either of these properties is expected to have confinement properties comparable to those of a tokamak. A good understanding of the nature of these particle confinement schemes is therefore very much needed. 
  
Guiding center theory for single particle motion demonstrates that radial drifts do not destroy confinement provided the second adiabatic invariant $\Jpl=\oint \vpl\:\dtl $ be independent of the magnetic field line label on a given flux surface. Thus omnigeneity can be defined as configurations where surfaces of constant $\Jpl$ are also flux surfaces. Quasisymmetry (QS) has been shown \cite{landremancatto2012omnigenity} to be a special case of omnigeneity. It requires more stringently that the strength of magnetic field be independent of the field line label in an appropriate coordinate system.
However attractive these ideas might be, there are some negative results in the literature regarding these optimization schemes. According to Garren\cite{garrenBoozer1991existence} and Boozer, the QS constraint can be satisfied exactly only on one flux surface and not in a given volume. According to Cary-Shasharina\cite{caryshashrinaPRL1997helical,caryshasharina1997omnigenity} an analytic omnigeneous magnetic field has to be QS. Until very recently \cite{parraLandreman2015lessconstrained}, it was also assumed  \cite{caryshasharina1997omnigenity,landremancatto2012omnigenity} that omnigeneous magnetic fields do not allow magnetic wells of varying depth. These results clearly indicate how stringent the constraints are and motivates further exploration. However, analytical progress is hindered by the inherent complications arising from the lack of adequate representations of the intrinsically three dimensional equilibrium.

In this paper, we study analytically the general question of radial localization of trapped particles in a nonsymmetric magnetohydrodyanmic (MHD) equilibrium. Our main objective is to obtain an analytical understanding of the constraints arising from omnigeneity and QS. We begin by analyzing the second adiabatic integral as an Abel integral as noted earlier by Hall-McNamara \cite{hall1975three}. In section \ref{sec:abelJpl}, we derive the well established results on omnigeneity through the explicit solution of an Abel integral equation. Abel inversion technique has been discussed in details in many excellent textbooks\cite{whittaker_watson_1996,bocher1914introduction,bracewell1986fourier} and articles \cite{keller_inv_problem}. Using Abel inversion one solves the inverse problem of determining a potential given the time period of a particle as a function of energy in that potential. Since the $\Jpl$ integral is of the form of an Abel integral \cite{hall1975three} we show that such an inversion is possible to obtain the magnetic well width measured along a field line. We present the Abel inversion results for a single and multiple wells in (\ref{subsec:abelJplsingle},\ref{subsec:abelJplmult}) respectively, analyzing the possibility of having an omnigeneous magnetic field with variable well depths in \ref{sec:sepa}.

Next, we turn to the question of evaluating the aforementioned constraints on an actual nonsymmetric MHD equilibrium. In section \ref{sec:3Deqb}, we discuss the basic framework to solve for three-dimensional MHD equilibrium based on an extension of the formulation developed earlier by Weitzner\cite{weitzner2016expansions}. We look for a local equilibrium expansion around the minima of a magnetic well in section \ref{sec:localexp}. The assumption of locality not only simplifies the analysis from  the geometrical point of view but also allows a study the physics of the deeply trapped particles. 

In sections \ref{sec:omniconstraint} and \ref{sec:QSconstraint}, we evaluate the omnigeneity and QS constraints and compare them in \ref{sec:comparison}. The constraints manifest themselves as algebraic equations that connect various coefficients in the local expansion of the magnetic field strength. We find that there is a redundancy in the constraint equations and the actual number of independent equations to be satisfied is significantly smaller than what is implied by a simple count of the number of constraints. We also find that within the context of a local analytic expansion, omnigeneity and QS are significantly different. The QS constraint starts at a lower order in the local expansion scheme and at each order it is more restrictive than the omnigeneity condition. Finally, we summarize our findings and discuss future directions in section \ref{sec:sum}.

\section{Reconstruction of $|\B|$ given $\Jpl$ using Abel integral inversion}
\label{sec:abelJpl}
Let us consider the motion of a particle of charge $e$ and mass $m$, trapped in a strong magnetic field $\B$ and an electrostatic potential $\varphi$ in a typical stellarator. The larmor radius is typically orders of magnitude smaller than the system size length. All of the particle confinement schemes are based on the lowest order guiding center description of such trapped particles and ignore any finite larmor radius effects. In the zero Larmor radius limit, the particle sees an effective potential given by $V= \mu B+(e/m)\varphi$, $\mu$ being the magnetic moment and $B=|\B|$ is the strength of the magnetic field. One can use the conservation \cite{helander2014theory} of the longitudinal invariant, $\Jpl$, given by an integral of the form 
\begin{align*}
\Jpl= \oint \vpl \:\dt s = \oint \sqrt{2(E-\mu B(s)-(e/m)\varphi)}\dt s,
\end{align*}
where $E$ is the total energy of the particle. The integral is done along a field line on a given flux surface ($\gamma =$constant) and a given field line label ($\Psi=$ constant) and $s$ measures length along $\B$. In equilibrium, the electrostatic potential is a flux function i.e $\varphi=\varphi(\gamma)$, which does not change along the path of the integral and therefore can be absorbed in the total energy $E$. This allows us to define a normalized kinetic energy $\cE$ so that the magnetic moment $\mu$ and the electrostatic potential do not explicitly appear in $\Jpl$ resulting in:
\begin{align}
\Jpl(\cE,\Psi,\gamma)= \oint_{\Psi,\gamma}\sqrt{2(\cE-B(s))}\dt s . \label{Jpl}
\end{align}
The bounce time $\tau$ can be formally obtained by differentiating $\Jpl$ with respect to $\cE$
\begin{equation}
\tau(\cE,\Psi,\gamma)=\frac{\partial \Jpl}{\partial \cE} =\oint_{\Psi,\gamma} \frac{\dt s}{\sqrt{2(\cE-B(s))}}.
\label{tauE}
\end{equation}
Provided the derivative of $\Jpl$ with respect to $\cE$ is continuous, the above equation is formally of an Abel type and we can use the Abel inversion machinery to obtain $s(B)$. However, continuity and differentiability of $\Jpl$ as a function of $\cE$ is a suspect near bounce points. Furthermore, in a multiple magnetic well system, there is a discontinuous jump in bounce points across separatrices. Therefore, we proceed to analyze first the single well and then the multiple well system.

\subsection{Abel inversion in a single magnetic well system}
\label{subsec:abelJplsingle}
 Let us consider the bounce points $\lbr \smpi{-}{}, \smpi{+}{}\rbr$ where $\cE=B$. In between the bounce points we have $\smpi{\text{min}}{}$ where the magnetic well has its minima $B_{\text{min}}$. In evaluating $\Jpl$ we split the region of integration into $\lbr \smpi{-}{},\smpi{\text{min}}{}\rbr \cup \lbr \smpi{\text{min}}{}, \smpi{+}{}\rbr$. At this stage it is useful to change integration variable from $s$ to $B$ to obtain
 \begin{align}
\Jpl = \int_{B_{min}}^{\cE} \dt B\:  \sqrt{2(\cE-B)}\cD ,
\label{Jplsingle}
 \end{align}
where 

\begin{align}
\cD=\sum_\sigma\left|\frac{ds}{dB}\right|_\pm =\sum_{\sigma=\pm 1} \at{\frac{1}{|\dpl B|}}{\pm}
,\:\: \sigma=\text{sign}\lbr\frac{ds}{dB}\rbr
\label{cD}
\end{align} 

The subscript $\pm$ is used to indicate that the quantity is evaluated at the bounce points. The sum is done at the bounce points where $\cE=B$. The discrete variable $\sigma=\pm 1$ characterizes the bounce points.

To investigate the behavior near a bounce point, let us consider a series expansion of $B(s)$ around the minima at $s=0$
\begin{align}
B=\Bmin +\frac{s^2}{2} B'' + O(s^3)\:,\:\: B''>0. \label{Bw}
\end{align}
To this order of approximation the system behaves like a simple harmonic oscillator with $\Jpl \propto \cE$ and $\tau \approx$ constant. Thus the $\Jpl,\tau$ integrals are convergent, differentiable and well behaved.

At the minimum $\cE=\Bmin$, the right side of (\ref{Jplsingle}) vanishes. The left side is also zero since the bounce points coalesce at the minima. We can therefore define the bounce function $\tau(\cE)$ as
\begin{align}
\tau(\cE,\Psi,\gamma)\equiv\frac{\partial \Jpl}{\partial \cE}=\int_{B_{min}}^\cE \frac{\dt B}{\sqrt{2(\cE-B)}} \cD ,\label{tausingle}
\end{align}
which is itself a continuous and differentiable function of $\cE$. Following the standard Abel integral procedure as outlined in Appendix \ref{appendix 0}, we obtain
 \begin{align}
 \cD(B)= \frac{2}{\pi}\frac{d}{dB} \int_{B_{\text{min}}}^B \dt\cE \frac{1}{\sqrt{2(B-\cE)}} \frac{\partial \Jpl}{\partial \cE} .
 \label{wsingle}
 \end{align}
Abel inversion guarantees continuity of $\cD$. Since the left side of (\ref{wsingle}) is a derivative one can integrate $\cD$ with respect to $B$. This gives us the width of the well equal to $\smpi{+}{}-\smpi{-}{}$ as can be checked from (\ref{cD}).
 \subsection{Abel inversion in a multiple magnetic well system}
 \label{subsec:abelJplmult}

\begin{figure}
\includegraphics[height=7cm]{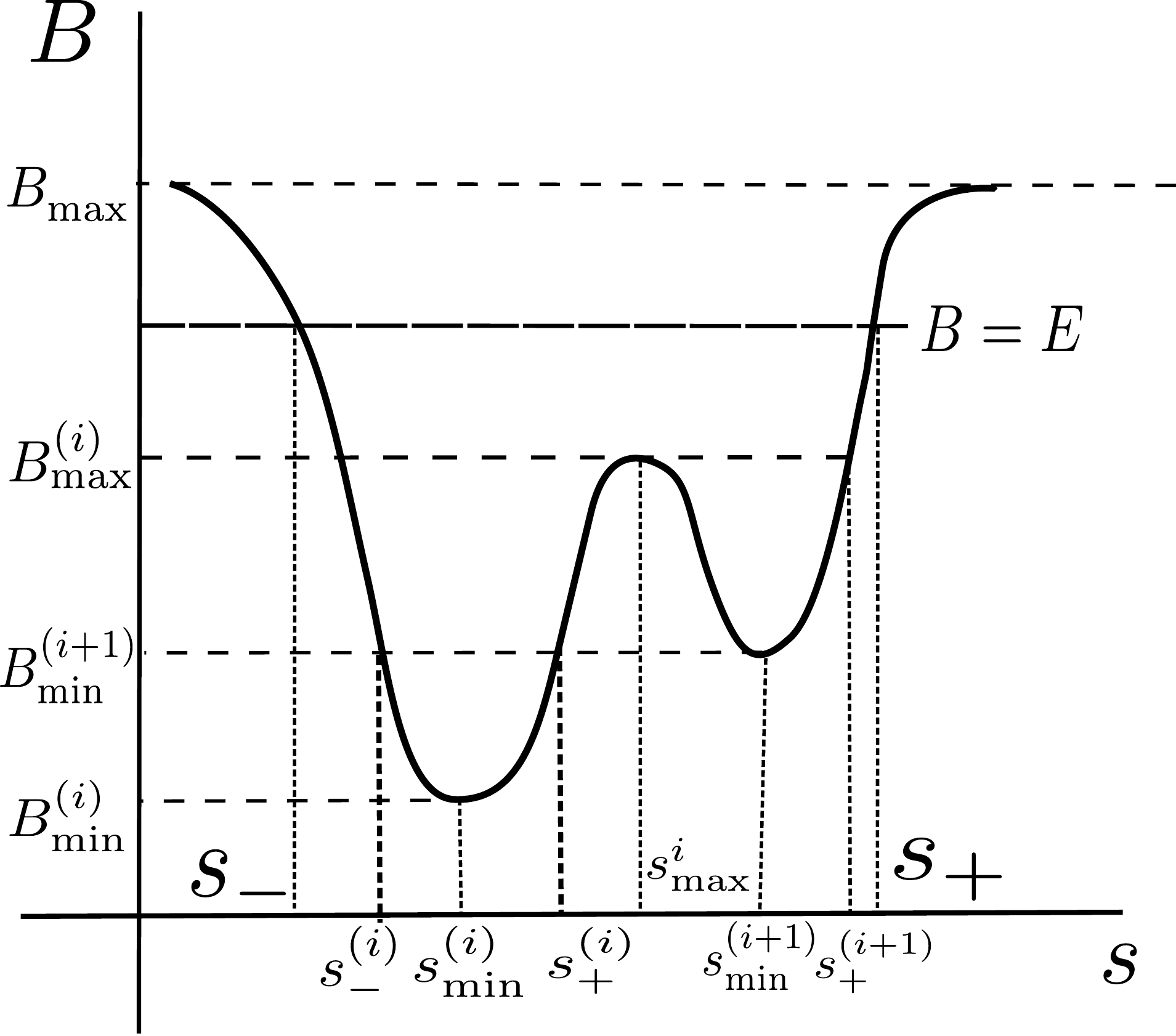}
\caption{Multiple magnetic well}
\label{fig1}
\end{figure}

Let us now consider a generic non-monotonic magnetic well with multiple wells as shown in Fig \ref{fig1}. For simplicity, we consider only two wells, however, the final results are completely general. Let us consider a particle with sufficient energy so that it can bounce back and forth between the two wells. We shall use the following notation: $(\smpi{-}{},\smpi{+}{})$ denotes the bounce points with $B=\cE$, $(\smpi{\text{min}}{(i)},\smpi{\text{max}}{(i)})$ denotes the location of the ith minima/ maxima $B=(B^{(i)}_{\text{min}},B^{(i)}_{\text{max}})$, $(\smpi{-}{(i)},\smpi{+}{(i)})$ denotes the locations where $B=B^{(i)}_{\text{max}}$ adjacent to the local maxima $B^{(i)}_{\text{max}}$. In evaluating $\Jpl$, we shall split the interval $(\smpi{\text{min}}{(i)},\smpi{\text{max}}{(i)})$ into the following sub intervals,
\begin{align*}
\lbr\smpi{-}{},\smpi{-}{(i)}\rbr\cup \lbr\smpi{-}{(i)},\smpi{\text{min}}{(i)} \rbr \cup\\ \lbr\smpi{\text{min}}{(i)},\smpi{+}{(i)}=\smpi{\text{max}}{(i)} \rbr \cup 
\lbr \smpi{\text{max}}{(i)}=\smpi{-}{(i+1)},\smpi{\text{min}}{(i+1)}\rbr  \cup  \\ \lbr\smpi{\text{min}}{(i+1)},\smpi{+}{(i+1)}\rbr  \cup \lbr \smpi{+}{(i+1)},\smpi{+}{} \rbr.
\end{align*}

Since, near the separatrix the bounce points vary discontinuously, we define a modified $\Jpl$ by subtracting out from $\Jpl$ the sum of the individual contributions from each well:

\begin{align*}
    J(\cE)=& \Jpl(\cE) -\sum_{j=i}^{i+1} \int_{B^{(j)}_{min}}^{B_{max}^{(i)}}\dt \cB \sqrt{2(\cE-\cB)\:}\cD^{(j)}\\
    =& \int_{B^{(i)}_{max}}^\cE \dt \cB\sqrt{2(\cE-\cB)}\: \cD .
\end{align*}
This is done to ensure that at $\cE=\Bmax^{(i)}$, $J$ vanishes just like $\Jpl$ vanished at the minimum in a single well. By restricting the inversion on the $J$ integral we avoid all points where $1/B'$ is unbounded except for the maxima at $s^{(i)}_{\text{max}}$. To ensure continuity at this point and around the separatrix, we shall assume the functions being subtracted are multiplied with suitable mollifiers. 

Now, let us test differentiability. The only points  where the function $J(\cE)$ could be non-differentiable are the bounce points near the separatrix. Let us consider the integral between bounce points $(\smpi{-}{},\smpi{+}{})$ where $\cE=B$. The integral with $s\geq \smpi{-}{}$ is $$J=\int_{\smpi{-}{}}^{s} \dt s\sqrt{2(\cE-B)}= \int_{\cE}^{B} \dt \cB\sqrt{2(\cE-\cB)}\:\frac{ds}{d\cB} $$
 
Let us now consider a series expansion of $B(s)$ around the maxima at $s=0$ (choosing $s^{(i)}_{\text{max}}=0$)
\begin{align}
B=\Bmax -\frac{s^2}{2} B'' +\frac{s^3}{3!} B'''+ O(s^4)\:,\:\: B''>0. \label{Bs}
\end{align}
From $B'(s)= -s B''+O(s^2)$ we obtain $s^2 \approx (\Bmax-B)/B''$ and therefore
$$J \simeq  \int_\cE^{B} \frac{\dt \cB}{\sqrt{\Bmax-\cB}}\sqrt{(\Bmax-\cB)-(\Bmax-\cE)}.$$ Changing variables from $\cB$ to $t=(\Bmax-\cE)/(\Bmax-B)$ we can simplify the J integral to be approximately \begin{align}
J\propto \int_1^{1-x} \sqrt{1-t}\approx x^{3/2} \label{Jsepa}
\end{align}
 with $x=1-t\sim(\cE -B)$ in the upper limit, being only slightly different from zero. This integral and its derivative is convergent and well behaved as required.
 
Now, that we have ensured continuity and differentiability of $J$ we can differentiate $J$ with respect to $\cE$ to obtain
\begin{align}
\frac{\partial J}{\partial \cE}&=\int_{\Bmax^{(i)}}^\cE \dt\cB\frac{1}{\sqrt{2(\cE-\cB)}}\: \cD\nonumber\\
&= \frac{\partial \Jpl}{\partial \cE}-\sum_{j=i}^{i+1} \int_{\Bmin^{(j)}}^{\Bmax^{(i)}}\dt\cB\frac{1}{\sqrt{2(\cE-\cB)}}\: \cD^{(j)} ,
\label{tauE2well}
\end{align}
 where $\cD^{(j)}$ is $\cD$ for the $j^{th}$ well. The first term in (\ref{tauE2well}) is analogous to Eq. (\ref{tauE}) although the lower bound is the local maxima instead of the minima as in (\ref{tauE}). The second term in (\ref{tauE2well}) is the sum of the contributions to the bounce time from the individual wells. Following the standard Abel integral inversion, the details of which are described briefly in Appendix \ref{appendix 0}, we obtain 

 \begin{align}
 &\cD=\frac{2}{\pi} \frac{d}{dB}\int_{\Bmax^{(i)}}^B \dt\cE \frac{1}{\sqrt{2(B-\cE)}}\frac{\partial \Jpl}{\partial \cE}+\label{wdouble}\\
 &\frac{2}{\pi}\frac{d}{dB} \sum_{j=1}^{i+1}\int_{\Bmin^{(j)}}^{\Bmax^{(i)}} \dt\cB\:\cD^{(j)}\: \text{sin}^{-1}{\sqrt{\frac{\Bmax^{(i)}-\cB}{B-\cB}}} .\nonumber 
 \end{align}
In the limit of a single well, $\Bmax^{(i)} \rightarrow \Bmin^{(j)}$ for all j,(\ref{wdouble}) implies that the second term is zero, and we recover the single well result (\ref{wsingle}). The second term is therefore a contribution from all the local magnetic wells. We can calculate $w=\smpi{+}{}-\smpi{-}{}$ by integrating (\ref{wdouble}) with respect to $B$.

\subsection{Abel inversion and omnigeneity}
\label{sec:sepa}
In an omnigeneous system, $\Jpl$ is independent of the field line label $\Psi$. Abel inversion allows us to calculate $\cD$ in terms of $\Jpl$. Therefore, in an omnigeneous system, $\cD$ must also be independent of $\Psi$ and the omnigeneity constraint can be written as
\begin{align}
\del_\Psi \cD = \del_\Psi \sum_{\sigma=\pm 1} \at{\frac{1}{|\dpl B|}}{\pm}=0.
\label{omnicrit0}
\end{align} 
 In the literature \cite{helander2014theory,landremancatto2012omnigenity}, this is also known as the ``Cary-Shasharina theorem". Note that this statement is equivalent to the requirement that the distance between the bounce points $w=\smpi{+}{}-\smpi{-}{}$, be independent of the field line label.

Let us now ask if the ``Cary-Shasharina theorem" (\ref{omnicrit0}) is valid for magnetic wells with varying well depths. It is well known that the bounce time of a particle diverges logarithmically near a separatrix.
 Expanding $B$ slightly below a local maxima (at $s=0$), as given by (\ref{Bs}) and using (\ref{tauE2well}), we get
 \begin{subequations}
 \begin{align}
 w= \wmax-\sqrt{2(\Bmax-B)/B''} \label{wsepa}\\
 \cD=\frac{dw}{d\cB} \simeq +\frac{1}{\sqrt{2B''(\Bmax-\cB)}} \label{dwsepa}\\
 \tau(\cE) =\frac{\partial \Jpl}{\partial \cE}\simeq\frac{1}{2\sqrt{B''}}\log|\cE-\Bmax| , \label{tausepa}
 \end{align}
 \end{subequations}
 where $\wmax$ is the well width at the local maxima. We recover the same scaling of the well width from the Abel inversion formulas (\ref{wsingle}-\ref{wdouble}), as can be seen by substituting (\ref{tausepa}) into (\ref{wdouble}) and evaluating the $\cE$ and $B$ integrals. 
 
It might seem from (\ref{dwsepa}) that the omnigeneity constraint is singular near a separatrix since $\cD$ diverges. But this is misleading since we can also interpret Cary-Shasharina theorem as the requirement that the distance between the bounce points $\smpi{+}{}-\smpi{-}{}$ be independent of the field line label. This distance approaches a finite value near the separatrix and hence the omnigeneity criteria still holds near a separatrix.

Previously \cite{caryshasharina1997omnigenity,landremancatto2012omnigenity}, it was assumed that having magnetic wells of different depths would imply lack of omnigeneity. The argument was based on the fact that near the separatrix the bounce time diverges logarithmically and a trapped particle on an average gains a large radial step size within one bounce time. Therefore, it was assumed that for omnigeneity there should be just one group of trapped particles. This restriction was later relaxed \cite{parraLandreman2015lessconstrained}. Our results are in agreement with these recent\cite{parraLandreman2015lessconstrained} clarifications.

\section{3D MHD equilibrium formulation}
\label{sec:3Deqb}

We begin our study of local 3D magnetohydrodynamic equilibrium to understand deeply trapped particle localization. In this section, we shall briefly review the basic nonsymmetric MHD equilibrium formulation as discussed in Weitzner \cite{weitzner2016expansions} and collect all the relevant equations. We recall that the representation applies generally in the neighbourhood of any point and may be extended into the full equilibrium domain.

The equilibrium is determined from the fundamental ideal MHD equations
\begin{align}
\J \times \B = \dl p, \quad \J= \dl \times \B, \quad \dl \cdot \B =0. \label{MHD}
\end{align}
We shall assume that the 3 D magnetic field $\B$ possesses nested flux surfaces $\gamma(x,y,z)$. A general coordinate system $(\alpha,\beta,\gamma)$ with $\alpha(x,y,z)$ and $\beta(x,y,z)$ denoting the ``poloidal" and ``toroidal" angles, can be constructed such that
\begin{align}
\Jac= \dl \alpha \times \dl \beta \cdot \dl \gamma = \frac{\partial(\alpha,\beta,\gamma)}{\partial(x,y,z)}>0.
\label{Jacdef}
\end{align}
 Pressure is assumed to be only a function of $\gamma$ , i.e, $p=p(\gamma)$ as in the axisymmetric case. We shall make use of the Grad-Boozer as well as the Clebsch representation for the magnetic fields
\begin{align}
\B= \dl \Phi + \zeta \dl \gamma, \quad \B= \dl \Psi \times \dl \gamma. \label{Bdef}
\end{align} 
$(\Phi,\Psi)$ can be thought of as components of scalar and vector potential for $\B$. They are harmonic conjugates when the current $\J$ is identically zero. From the Grad-Boozer representation of $\B$ we obtain current $\J$ by taking the curl of $\B$ and $\zeta$ by dotting with $\dl \alpha\times \dl \beta$. Thus,
\begin{subequations}
\begin{align}
    \J &= \dl \zeta \times \dl \gamma \label{Jdef}\\
    \zeta &= -\Phi_{,\gamma}+\Jac^{-1}\B\cdot \lbr \dl \alpha \times \dl \beta \rbr.  \label{zetdef}
\end{align}
\end{subequations}
Alternatively, using the fact that $\B\cdot \dl p =\J\cdot \dl p=0$ we obtain,
\begin{subequations}
\begin{align}
\zeta &= -(\dl \Phi\cdot \dl \gamma)/|\dl \gamma|^2 \label{altzetdef}\\
\B &= \dl \gamma \times (\dl \Phi\times \dl \gamma)/|\dl \gamma|^2 . \label{Bfidef}
\end{align}
\end{subequations}
Note that Eq.(\ref{Bfidef}) implies
\begin{align}
\B\cdot\dl \Phi=B^2. \label{Bdfi}
\end{align}
We can relate $(\Phi,\Psi)$ by equating the two expressions for $\B$ as given in (\ref{Bdef},\ref{Bfidef}) to obtain
\begin{align}
\dl \Psi \times \dl \gamma =  \dl \gamma \times (\dl \Phi\times \dl \gamma)/|\dl \gamma|^2. \label{sifi}
\end{align}
Taking components of (\ref{sifi}) with respect to $(\alpha,\beta)$  results in
\begin{align}
\begin{pmatrix}
\Phi_{,\alpha}\\ \Phi_{,\beta}
\end{pmatrix}=\Jac
\begin{pmatrix}
g_{\alpha\alpha} \quad -g_{\alpha\beta}\\ 
g_{\alpha\beta} \quad -g_{\beta\beta}
\end{pmatrix}
\begin{pmatrix}
\Psi_{,\beta}\\ \Psi_{,\alpha}
\end{pmatrix},
\label{sifimatrix}
\end{align}
where
\begin{subequations}
\begin{align}
g_{\alpha\alpha}&=\frac{1}{\Jac^{2}}|\dl\beta\times\dl \gamma|^2,\:\: g_{\beta\beta}=\frac{1}{\Jac^{2}}|\dl\gamma\times\dl\alpha |^2 \\
g_{\alpha\beta}&=\frac{1}{\Jac^{2}}(\dl\beta\times\dl \gamma)\cdot (\dl\gamma\times\dl\alpha )\label{gmetric}\\
g &=g_{\alpha\alpha}g_{\beta\beta}-g_{\alpha\beta}^2=\frac{|\dl \gamma|^2}{\Jac^2}
\end{align}
\end{subequations}
are the metric elements of the magnetic flux surface.

Finally, substituting the expression of $\J$ from (\ref{Jdef}) into the equilibrium equation (\ref{MHD}) we obtain 
the nonsymmetric generalized ``Grad-Shafranov" equation
\begin{align}
\B\cdot \dl \zeta= p'(\gamma), \quad \B\cdot \dl =\Jac(\Psi_{,\beta}\partial_\alpha - \Psi_{,\alpha}\partial_\beta ), \label{GS0}
\end{align}
where $\zeta$ can be obtained either from (\ref{zetdef}) or (\ref{altzetdef}).

 In the following, we shall specialize to a topological torus with $\alpha=y, \beta =z$. This would simplify the equations considerably while still retaining the major difficulties associated with finding a generic nonsymmetric equilibrium.
 
\subsection{Equilibrium formulation I}
 Following Weitzner\cite{weitzner2014ideal,weitzner2016expansions}, we shall do an inversion from $\gamma(x,y,z)$ to $x(y,z;\gamma)$. Note that, $\gamma$ is now being treated as a parameter and the angle derivatives will now be at fixed $\gamma$. In this system, the metric coefficients are given by
\begin{subequations}
\begin{align}
\Jac^{-1}=x_{,\gamma}, \quad g_{\alpha\beta}=x_{,y}x_{,z}\\
g_{\alpha\alpha}=1+x_{,y}^2 ,\quad g_{\beta\beta}=1+x_{,z}^2\\
g= x_{,\gamma}^2|\dl \gamma|^2 = 1+x_{,y}^2+x_{,z}^2.
\end{align}
\end{subequations}
Thus, Eq.(\ref{sifimatrix}) and (\ref{zetdef}) yields
\begin{align}
x_{,\gamma}\begin{pmatrix}
\Phi_{,y}\\ \Phi_{,z}
\end{pmatrix}=
\begin{pmatrix}
1+x_{,y}^2 \quad -x_{,y}x_{,z}\\ 
\quad x_{,y}x_{,z} \quad -(1+x_{,z}^2)\quad
\end{pmatrix}
\begin{pmatrix}
\Psi_{,z}\\ \Psi_{,y}
\end{pmatrix}
\label{sifixy}
\end{align}
and
\begin{subequations}
\begin{align}
-\zeta= \Phi_{,\gamma}+(x_{,z}\Psi_{,y}-x_{,y}\Psi_{,z}) \label{zetaI}\\
 x_{,\gamma}\B\cdot \dl =(\Psi_{,z}\partial_y - \Psi_{,y}\partial_z).
\label{BDI}
\end{align}
\end{subequations}
The ``Grad-Shafranov" (\ref{GS0}) equation can be shown to be
\begin{align}
-x_{,\gamma}p'(\gamma)=\left(\Psi_{,z}\partial_y - \Psi_{,y}\partial_z\right)\left( \Phi_{,\gamma}+(x_{,z}\Psi_{,y}-x_{,y}\Psi_{,z})\right). \label{GSI}
\end{align}

\subsection{Equilibrium formulation II}
Alternatively, we could do a double inversion and treat both $\gamma$ and $\Psi$ as parameters. We shall regard $x,y,\Phi$ as functions of $(\gamma,\Psi,z)$. It can be shown that
\begin{align*}
(\Psi_{,x},\Psi_{,y},\Psi_{,z})=\Jac \left(\ycg,-\xcg,(\xcg\:\ycz -\ycg\:\xcz)\right) \nonumber\\
(\gamma_{,x},\gamma_{,y},\gamma_{,z})=\Jac \left(-\ycsi,\xcsi,(\xcz\:\ycsi -\ycz\:\xcsi)\right) ,
\end{align*}
where the Jacobian $\Jac$ is given by,
\begin{align}
\Jac^{-1}=\{x,y\}_{(\Psi,\gamma)}= x_{,\Psi}y_{,\gamma}-y_{,\Psi}x_{,\gamma} \label{JacII}.
\end{align}
The main equations in this formulation are given by,
 \begin{subequations}
 \begin{align}
 \Phi_{,z}&= \Jac\:(1+x_{,z}^2+y_{,z}^2) \label{fiz}\\
 \Phi_{,\Psi}&= \Jac\:(x_{,z}x_{,\Psi}+y_{,z}y_{,\Psi})\label{fisi}\\
 \zeta +\Phi_{,\gamma}&= \Jac\:(x_{,z}x_{,\gamma}+ y_{,z}y_{,\gamma}).
 \label{zetaII}
 \end{align}
 \end{subequations}

The expression of $\B\cdot \dl$ is particularly simple in this formalism:
\begin{align}
\B\cdot \dl = \Jac \partial_z .\label{BDII}
\end{align}

The non-axisymmetric ``Grad-Shafranov" equation is now given by
\begin{align}
-p'(\gamma)\Jac^{-1}=\partial_z \left( \Phi_{,\gamma} -\Jac(x_{,z}x_{,\gamma}+y_{,z}y_{,\gamma}) \right). \label{GSII}
\end{align}

We can eliminate $\Phi$ completely using Eqs.(\ref{fiz}), (\ref{fisi}) and (\ref{zetaII}) to obtain a coupled set of equations for $(x,y)$ as functions of $(\Psi,\gamma,z)$
\begin{subequations}
\begin{align}
\partial_\psi(\Jac(1+x_{,z}^2+y_{,z}^2))&= \partial_z( \Jac\:(x_{,z}x_{,\Psi}+y_{,z}y_{,\Psi})) \label{xyeqn}\\
\partial_\gamma \left(\Jac(1+x_{,z}^2+y_{,z}^2)\right)&= \partial_z \left( \Jac(x_{,z}x_{,\gamma}+y_{,z}y_{,\gamma}) \right)\label{GSxy}\\
& \quad -p'(\gamma)\Jac^{-1} . \nonumber
\end{align}
\end{subequations}

The magnetic field and its strength are given by
\begin{subequations}
\begin{align}
\B &= \Jac\:(\zh + \xcz\: \xh + \ycz\:\yh)\label{BII}\\
B&= \Jac(1+\xcz^2+\ycz^2)^{1/2}. \label{BsqII}
\end{align}
\end{subequations}

In this formulation for fixed $(\gamma,\Psi)$, $z$ varies along field lines. Thus, this is particularly suited to describe trapped particles, which to lowest order in Larmor radius expansion, follow field lines between successive bounces. In the following we shall use this formalism to obtain a local equilibrium expansion. 

\section{Local non-symmetric equilibrium expansion}
\label{sec:localexp}
We are ultimately interested in studying the confinement of deeply trapped particles in a nonsymmetric MHD equilibrium and therefore, a local expansion around a local minima of the magnetic well is sufficient. Further the results apply whether or not the field lines are ergodic and flux surfaces exist. As the expansion is local about a point, a power series expansion in the variables $(\Psi,\gamma,z)$ is adequate.
 We seek expansions of the generic form \begin{align*}
     A(\Psi,\gamma,z)&=\sum_n \ep^n A^{(n)}(\Psi,\gamma,z)\\
     A^{(n)}(\Psi,\gamma,z)&=\sum_{i+j+k=n} A_{ijk}\gamma^i \Psi^j z^k,
\end{align*} 
 where $A_{ijk}$ are constant coefficients and $\epsilon$ has been inserted to keep track of the terms. Expanding the relevant quantities about $(0,0,0)$, we obtain
\begin{subequations}
\begin{align}
\ep x(\Psi,\gamma,z)=& \ep \gamma +\epsilon^2(x_{200} \gamma^2 + x_{110} \gamma \Psi+ x_{101} \gamma z \nonumber\\
&+ x_{011} z \Psi+ x_{020} \Psi^2 +x_{002} z^2)+ O(\epsilon^3) \label{xser}\\
\ep y(\Psi,\gamma,z)=& -\ep \Psi + \epsilon^2 (y_{200} \gamma^2 + y_{110} \gamma \Psi+ y_{101} \gamma z \nonumber\\
&+ y_{011} z \Psi+y_{020} \Psi^2 +y_{002} z^2)+ O(\epsilon^3) \label{yser}\\
\ep\Phi(\Psi,\gamma,z)=& \ep z + \epsilon^2 (\Phi_{200} \gamma^2 + \Phi_{110} \gamma \Psi+ \Phi_{101} \gamma z \nonumber\\
&+ \Phi_{011} z \Psi+y_{020} \Psi^2 +\Phi_{002} z^2)+ O(\epsilon^3)\\
p(\gamma)=&p_0 +\epsilon p_1 \gamma + \epsilon^2 p_2 \gamma^2 +O(\epsilon^3). \label{phiser}
\end{align}
\end{subequations}
 
 The lowest order terms are normalized such that $x=\gamma$ denotes a flux surface, $y=-\Psi$ denotes a field line on $\gamma$ and the lowest order magnetic field is given by $\B= \zh$. Using these expansions, we shall perturbatively solve the equation set (\ref{xyeqn}-\ref{GSxy}) around a local minima of the magnetic field strength along a field line. 
 
 Before proceeding to solve the MHD equilibrium equations order by order, we note that we can eliminate all terms of the form $(x_{n00}\gamma^n,y_{0n0}\Psi^n)$ by redefining $\gamma$ and $\Psi$ respectively. We can also eliminate terms of the form $\Phi_{n00}\gamma^n$ through a redefinition of $\zeta$, see (\ref{Bdef}). In order to understand which coefficients are independent we solve the MHD equations for the coefficients $(x_{ijk},\Phi_{ijk})$ in terms of $y_{ijk}$. Those not expressable in terms of the $y_{ijk}$ and the $y_{ijk}$ themselves are therefore independent. 
 
The magnetic field and its strength can be evaluated from (\ref{BII}-\ref{BsqII}). $|\B|$ thus obtained is of the form
\begin{align}
B(\Psi,\gamma,z)&= 1 +\epsilon (B_{100}\gamma +B_{010}\Psi +B_{001}z)\nonumber \\
&+\epsilon^2(B_{200} \gamma^2 + B_{110} \gamma \Psi + B_{101} \gamma z+ B_{011} z \Psi  \nonumber\\
&+ B_{020} \Psi^2 +B_{002} z^2)+ O(\epsilon^3), \label{Bexpan}
\end{align}
where the coefficients $B_{ijk}$ are polynomial functions of the coefficients $(x_{ijk},y_{ijk})$.

Next, we give a systematic analysis of the nonsymmetric equilibrium expansion calculation. The calculations are straightforward but tedious. We have found it useful to use Mathematica.

 To first order in $\epsilon$, expanding $B$ using (\ref{Bexpan},\ref{BsqII}) and $\Jac=1+O(\epsilon)$, we get
\begin{align*}
&B_{001}=(x_{101}-y_{011}),\:\: B_{100}=-\:y_{110},\:\:B_{010}= x_{110}.
\end{align*}
Using $\B\cdot\dl = \Jac \partial_z$, we obtain,
$$\dpl B \propto \epsilon B_{001}+O(\ep^2).$$
Thus, $B_{001}=0$ i.e $x_{101}=\:y_{011}$ so that $z=0$ is a minimum along $\B$. From (\ref{fiz}) we get $\Phi_{101}=y_{110}$ so that $B$ is 
\begin{align}
B= 1+ \ep(B_{100} \gamma +B_{010}\Psi)+O(\epsilon^2) \nonumber\\
\text{with} \quad B_{100}=- \gamma\: y_{110}, \quad B_{010}=- 2\: y_{002}.
\label{Bsqrseries}
\end{align}
The parallel derivative vanishes to first order by construction. To second order it is given by
\begin{align}
\dpl B =\ep^2(B_{101}\gamma + B_{011}\Psi +2 B_{002} z)+O(\epsilon^3). \label{dplb}
\end{align}
The minimum at $z=0$ occurs on a given filed line $\Psi=$ constant on the surface $\gamma=$ constant and is not a global minima.
Finally, equations (\ref{xyeqn},\ref{GSxy}) to first order, imply 
\begin{align}
x_{110}= 2 y_{002}, \:\: x_{002}= \frac{1}{2}(p_1+y_{110}).\nonumber
\end{align}
The $\Phi$ equations (\ref{fiz},\ref{fisi}) to this order yields
\begin{align}
&\Phi_{002} = 0, \quad \Phi_{011} = -2 y_{002},\:\: 2\Phi_{020} = -y_{011},\\
&\quad \Phi_{110} = -y_{101}, \quad \Phi_{101}= y_{110}. \nonumber
\end{align}
The independent coefficients to this order are two from $x_{ijk}$ and five $y_{ijk}$, namely $(x_{011},x_{020})$ and $\{y_{200},y_{110},y_{101},y_{011},y_{002}\}$. 

The expansions can be carried to higher order without any difficulty and we present the results in Appendix \ref{appendix B}. The next order independent coefficients are three x's $\{x_{030},x_{003},x_{021}\}$ and nine y's $\{y_{300},y_{003},y_{210},y_{201}, y_{120},y_{102},y_{111},y_{012},y_{021}\}$. Similarly we obtain $\{x_{301},x_{040},x_{031}\}$ as the three independent coefficients of fourth order.
 
 Now that we have a local MHD equilibrium expansion, we proceed to evaluate the omnigeneity and QS conditions in the next sections.

\section{Omnigeneity constraint }
\label{sec:omniconstraint}
The omnigeneity constraint, unlike QS, is not a local constraint. It is clear from the Cary-Shasharina theorem (\ref{omnicrit0}), that the constraint involves the parallel derivative of $B$ evaluated at the two bounce points of the trapped particles. Therefore, we first need to obtain expressions for the bounce points. The advantage of using the field line following coordinates $(\Psi,\gamma,z)$ now becomes obvious since, the bounce points are simply given by $z=\zb, B(\Psi,\gamma,\zb)=\Bb$ on a given field line ($\Psi=$ constant) and on a given flux surface ($\gamma=$ constant). We further expand $\zb$ and $\Bb$ in $\ep$  
\begin{subequations}
\begin{align}
\zb&=\zb_0 +\epsilon \: \zb_1+\epsilon^2 \: \zb_2 +..\label{zb}\\
\Bb&=\Bb_0 +\epsilon \: \Bb_1+\epsilon^2 \: \Bb_2 +..\label{Bb}
\end{align}
\end{subequations}

Using (\ref{zb},\ref{Bb}) and (\ref{Bexpan}) for $\Bb$, we can algebraically obtain the lowest order bounce point $z=\zb_0$ and higher order corrections $\zb_1,\zb_2$ etc. Note that $\Psi$ and $\gamma$ will enter $z$ and $B$ as parameters. Thus, the omnigeneity condition can be obtained by equating the $\Psi$ derivative of the quantity $\cD$ (as defined in (\ref{cD})) to zero. From (\ref{cD}), we observe that $\ep^2\cD \sim \ep^2 (\dpl B)^{-1}\sim O(1)$. 

In the next sections, we shall obtain expressions for the bounce point and some useful identities before calculating the omnigeneity constraint explicitly. 

\subsection{Calculation of bounce points}
Equating Eq. (\ref{Bb},\ref{Bexpan}) using (\ref{zb}), we obtain up to $O(\epsilon^2)$
\begin{subequations}
\begin{align}
&\Bb_0=1, \quad \Bb_1 = -2 \:\Psi y_{002}+\gamma\: y_{110}\\
&B_{002}\:\zb^2_0 + (\Psi\: B_{011}+\gamma\: B_{101})\:\zb_0 \label{z0eqn} \\
 &+(\gamma^2\: B_{200}+ \gamma\:\Psi\: B_{110}+\Psi^2\: B_{020} -\Bb_2)=0 \nonumber
\end{align}
\end{subequations}
Since Eq. (\ref{z0eqn}) is a quadratic equation for the lowest order bounce point $\zb_0$ we have two solutions $\zb_0^{+},\zb_0^{-}$ denoting the left and right bounce points. Using the discrete variable $\sigma\: (=\pm 1)$, as defined in Eq.(\ref{cD}) we have from (\ref{z0eqn})
\begin{subequations}
\begin{align}
\sum_{\sigma=\pm 1} \zbs_0 =\zbp_0 +\zbm_0=-\frac{1}{B_{002}}(\Psi\: B_{011}+\gamma\: B_{101})\label{z0ssum}\\
\zbp_0 \zbm_0=\frac{1}{B_{002}} (\gamma^2\: B_{200}+ \gamma\:\Psi\: B_{110}+\Psi^2\: B_{020} -\Bb_2)\label{z0pz0m}\\
\sigma |\zbp_0-\zbm_0|\:B_{002}= 2\:B_{002}\:\zbs_0 +B_{101}\:\gamma +B_{011}\:\Psi \label{deltaz0}.
\end{align}
\end{subequations}
Going to next order, we obtain an equation for $\zbs_1$
\begin{align}
\Bb_3 &= \lbr 2B_{002}\zbs_0 +\Psi\: B_{011}+\gamma\: B_{101}\rbr \zbs_1\nonumber\\
&+ B_{003}(\zbs_0)^3 + (\Psi\: B_{012}+\gamma B_{102})(\zbs_0)^2\label{z1s}\\ &+(B_{201}\gamma^2+B_{111}\gamma\Psi+B_{021}\Psi^2)\zbs_0 \nonumber\\
&+\lbr \gamma^3 B_{300}+B_{210}\gamma^2\:\Psi +B_{120}\gamma \Psi_2 +B_{030}\Psi^3 \rbr \nonumber
\end{align}

Using the bounce point equations for $\zb_0$ the following useful identity can be derived (details given in Appendix \ref{appendix C})
\begin{align}
&-|\Delta \zb_0|^2\lbr B_{002}\frac{\Delta \zb_1}{\Delta \zb_0} +(\Psi\: B_{012}+\gamma\: B_{101})\rbr  \label{z1diff}  \\
\begin{split}
&=\lbr \sum_\sigma \zbs_0\rbr
\left(\vphantom{\lbr\zbp_0+\zbm_0\rbr^2} (B_{201}\gamma^2+B_{111}\gamma\Psi+B_{021}\Psi^2)  \right. \\
& \quad \left. \quad \quad \quad +B_{003}\lbr \lbr\zbp_0+\zbm_0\rbr^2-3 \zbp_0 \zbm_0\rbr \right)
 \end{split} \nonumber\\
&\:  +2 \zbp_0 \zbm_0 \lbr B_{102}\gamma +B_{012}\Psi \rbr \nonumber \\
&\:+ 2\lbr-\Bb_3 +\gamma^3 B_{300}+B_{210}\gamma^2\:\Psi +B_{120}\gamma \Psi_2 +B_{030}\Psi^3 \rbr\nonumber
\end{align}
where $\Delta \zb_n = \zbp_n -\zbm_n$ for $n=0,1.$

\subsection{Omnigeneity criteria}
\label{subsec:omnicrit}
We are finally in a position to evaluate the omnigeneity criteria. Since we have expanded about a minima of the magnetic well, $\dpl B$ is $O(\epsilon^2)$. To $O(1)$
\begin{subequations}
\begin{align}
\ep^2\at{(\dpl B)^{-1}}{\pm}= &\frac{1}{\alpha^{(\sigma)}_1} +O(\epsilon^{1})\\
\ep^2\cD=\sum_{\sigma=\pm 1}\sigma\at{(\dpl B)^{-1}}{\pm}=& \sum_\sigma (\sigma\,\alpha^{\sigma}_1)^{-1} +O(\epsilon^{1}), \label{cDm2}
\end{align}
\end{subequations}
where
\begin{align}
\alpha_1^{(\sigma)}&= 2\:B_{002}\:\zbs_0 +B_{101}\:\gamma +B_{011}\:\Psi\\
&= \sigma |\zbp_0-\zbm_0|\:B_{002} =\sigma B_{002} \Delta \zb_0\nonumber
\end{align}
In the last equation we have used (\ref{deltaz0}). Summing over $\sigma$ we obtain
$$\ep^2\cD = \frac{2}{B_{002} \: \Delta \zb_0} + O(\epsilon^{1}).$$
We can calculate $\Delta\zb_0$ from (\ref{z0eqn}) and (\ref{z0ssum},\ref{z0pz0m})
\begin{align}
    \Delta\zb_0^2= &(\zbp+\zbm)^2-4\zbp\zbm \nonumber\\
    \partial_\psi \Delta\zb_0^2= &2\left(B_{011}^2 - 4 B_{002}B_{020} \right)\Psi \label{zb0psi}
    \\&-2\left(2 B_{002}B_{110}-B_{011}B_{101} \right)\gamma. \nonumber
\end{align}
Omnigeneity condition (\ref{omnicrit0}) requires $\cD$ to be independent of $\Psi$. Hence, to this order, the well width $\Delta \zb_0$ needs to be independent of $\Psi$ to ensure that the system is omnigeneous.
Since $\partial_\Psi \cD$ furnished us a linear polynomial in $(\gamma,\Psi)$ we expect two equations. From (\ref{zb0psi}) we indeed obtain two equations
\begin{align}
&4 B_{002}B_{020}=B_{011}^2\nonumber \\
&2 B_{002}B_{110}=B_{011}B_{101}.
\label{omniB1}
\end{align}
From Appendix \ref{appendix B}, we find that these equations translate into nonlinear low order polynomial equations for the two independent coefficients $\{x_{020},x_{011}\}$. 

To next order,
\begin{subequations}
\begin{align}
&\ep^2\at{(\dpl B)^{-1}}{\pm}= \frac{1}{\sigma\alpha_1}  -\frac{\alpha^{(\sigma)}_2}{\alpha_1^2}\ep+O(\epsilon^{2})\\
&\ep^2\cD=\ep^2\sum_{\sigma=\pm 1} \sigma \at{(\dpl B)^{-1}}{\pm}=\frac{2}{\alpha_1}-\frac{\epsilon}{\alpha_1^2} \sum_\sigma \sigma\:\alpha^{\sigma}_2 +O(\epsilon^{2}), \label{cDm1}
\end{align}
\end{subequations}
where
\begin{subequations}
\begin{align}
\alpha_1&=|\alpha^{(\sigma)}_1|= B_{002}|\zbp_0-\zbm_0|= B_{002}\Delta \zb_0 \label{alf1}\\
\alpha^{(\sigma)}_2&= 3 B_{003}\:(\zbp_0)^2+2\:B_{002}\:\zbs_1\label{alf2}\\
&\quad+2(\Psi\: B_{012}+\gamma\: B_{102})\zbs_0 \nonumber \\
&\quad+(\gamma^2\: B_{201}+ \gamma\:\Psi\: B_{111}+\Psi^2\: B_{021} -\Bb_2).
\nonumber
\end{align}
\end{subequations}

Using (\ref{z0ssum},\ref{z1diff}) and (\ref{alf1},\ref{alf2}), it can be shown (details in Appendix \ref{appendix C}) that
\begin{subequations}
\begin{align}
&\sum_\sigma \sigma\alpha^{(\sigma)}_2 = (\alpha^{(+)}_2-\alpha^{(-)}_2)= \label{alf2sum}\\
&\Delta \zb_0 \lbr 3 B_{003} \sum_\sigma \zb_0+2\lbr B_{002} \frac{\Delta \zb_1}{\Delta \zb_0}+\Psi\: B_{012} +\gamma\: B_{102}\rbr \rbr , \nonumber
\end{align}
\end{subequations} 
which can be evaluated using the expressions given in (\ref{z0ssum},\ref{z1diff}). Beyond this expression the calculation is laborious and is basically collecting terms with similar powers of $(\gamma,\Psi)$. Since the lowest order well width $\Delta\zb_0$ and $\alpha_1$ are independent of $\Psi$, the omnigeneity constraint, $\cD_\Psi=0$ is equivalent to the expression given in equation (\ref{alf2sum}) being independent of $\Psi$.

Evaluating $\partial_\Psi\cD=0$, we obtain a polynomial equation spanned by $\{1,\Psi^2,\Psi\gamma, \gamma^2\}$. Equating the coefficients of this quadratic polynomial to zero we get the following four equations that can be solved for the third order coefficients $\{B_{012},B_{030},B_{210},B_{120}\}$,

\begin{align}
 2 B_{002}\:B_{012} = &3 B_{003} B_{011} \nonumber\\
 4 B_{002}^3 B_{030} = &- B_{003} B_{011}^3 + 2 B_{002}^2 B_{011} B_{021})\label{omniB2}\\
 2 B_{002}^2 B_{210} = &B_{101} B_{111}B_{002} - B_{011} (B_{101} B_{102} -B_{002} B_{201}) \nonumber\\
 4 B_{002}^3 B_{120} = &-4 B_{002}^2( B_{021} B_{101}+ B_{011} B_{111})\nonumber\\ 
 &-B_{011}^2 ((3 B_{003} B_{101}) + 
 2 B_{002}B_{102}) .\nonumber
\end{align}

They determine the set $\{B_{012},B_{030},B_{210},B_{120}\}$. These equations can be used to determine the independent coefficients $\{x_{003},x_{030},y_{110},y_{101}\}$. Note that we still have many more independent coefficients left unconstrained by omnigeneity to this order. In the next section, we shall evaluate the QS constraints, and then compare them with the results of this section.

\section{Quasisymmetry constraint}
\label{sec:QSconstraint}
In the literature, there are many different forms of QS. The fundamental quasisymmetry requirement is that the parallel gradient of $|\B|$  on a given flux surface $\gamma$, must be a function of only $\gamma$ and $|\B|$, and independent of the field line label $\Psi$, i.e
\begin{align}
\dpl B = f(\gamma, B) \quad \Leftrightarrow \quad \partial'_\Psi(\dpl B) =0,
\label{QS0}
\end{align}
where the prime on $\partial'_\Psi$ indicates that $(\gamma,B)$ is held fixed instead of $(\gamma,z)$. Furthermore, if the MHD equilibrium equations (\ref{MHD}) together with the above constraint (\ref{QS0}) are satisfied, then it can be shown \cite{helander2014theory} that
\begin{align}
\B\cdot\dl \lbr \frac{\B \times \dl \gamma \cdot \dl B}{\B\cdot \dl B}  \rbr=0.
\label{BDBpsi}
\end{align}

For general ergodic field lines, (\ref{BDBpsi}) implies that 
\begin{align}
\frac{\B \times \dl \gamma \cdot \dl B}{\B\cdot \dl B} = B_\Psi(\gamma)
\label{Bpsidef}
\end{align}
i.e $B_\psi$ must be a function of $\gamma$ alone with
\begin{align}
\partial_\Psi B_\Psi =0, \quad \partial_z B_\Psi =0.
\label{QH}
\end{align}
Note that $B_\Psi$ is singular at the magnetic minima unless the numerator also vanishes simultaneously along with the denominator. This suggests that the function $B_\Psi$ can be discontinuous across the magnetic minima. However, in the case of ergodic field lines, we can continue on a given line that crosses the minima several times and since $B_\Psi$ can not vary across field lines (from (\ref{BDBpsi})), the only possible solution to (\ref{BDBpsi}) is (\ref{Bpsidef}). 

In our local analysis, there is no way of determining whether a field line is ergodic or not, and $B_\Psi$ can be $\Psi$ dependent. The constraints (\ref{BDBpsi}) and (\ref{QH}) are therefore different from a local analysis point of view. Since magnetic field lines in general are ergodic, (\ref{QH}) is the most common and relevant form of the QS constraint. We shall now explore the different forms of the QS constraints, (\ref{QS0},\ref{BDBpsi}) and (\ref{QH}), and compare them with omnigeneity constraints.

\subsection{QS as a special case of omnigeneity}

Comparing (\ref{QS0}) with the omnigeneity condition (\ref{omnicrit0}), we find that any QS system must be omnigeneous but there is no guarantee that an omnigeneous system will be QS.

We can express the partial derivative $\partial'_\Psi$ in terms of the usual $\partial_\Psi$ at fixed $(\gamma,z)$ through 
\be
\partial'_\Psi=\partial_\Psi+z'_\Psi\partial_z,
\ee
where $z'_\Psi=\partial'_\Psi z$. Before we can evaluate the QS constraints we need an expression for $z'_\Psi$. Operating $\partial'_\Psi$ on the series expansion of $B$ (Eq. \ref{Bexpan}), we obtain an equation for $z'_\Psi$
\begin{align}
B_{010} +\epsilon ((B_{110} \gamma +B_{011}z +2B_{020}\Psi)\nonumber \\
+\epsilon(B_{101}\gamma +B_{011}\Psi +2 B_{002}z)z'_\Psi +O(\ep^2)=0,  \label{zpsi}
\end{align}
which can be solved perturbatively to obtain $z'_\Psi$. Since $z'_\Psi \sim O(1)$, we find from the above equation that $B_{010}$ must vanish. Thus, we see that the lowest order QS constraint is given by 
\begin{align}
B_{010}=-2y_{002}=0.
\label{QSreq1}
\end{align}
Equations (\ref{QSreq1}) and (\ref{Bexpan}) show that quasisymmetry constrains even the first order term in $|\B|$. The series expansion for $z'_\Psi$ now takes the form $$z'_\Psi = (z'_\Psi)^{(0)}+(z'_\Psi)^{(1)}\ep +O(\ep^2),$$ which can be  solved to obtain 
\begin{align}
z'_{\Psi} = -\frac{2 B_{020}\Psi +B_{110}\gamma +B_{011}z}{B_{011}\Psi +B_{101}\gamma +2B_{002}z}+O(\ep).
\label{zpsi0}
\end{align}

Let us now evaluate the QS constraint (\ref{QS0}) order by order. From (\ref{dplb}), we find that to lowest order the $\Psi$ and $z$ derivatives of $\dpl B$ are given by $\ep^2(B_{011},2B_{002})$ respectively. The QS constraint (\ref{QS0}) therefore takes the form
\begin{align}
B_{011}+2B_{002}(z'_\Psi)=0.
\label{zpsi0eq}
\end{align}

From (\ref{zpsi0eq}) we find that $z'_\Psi$ should be a constant to lowest order. Substituting the expression of $z'_\Psi$ from (\ref{zpsi0}) into equation (\ref{QSreq1}) we obtain to lowest order
\begin{align}
(z'_\Psi)^{(0)}=-\frac{2 B_{020}\Psi +B_{110}\gamma +B_{011}z}{B_{011}\Psi +B_{101}\gamma +2B_{002}z}=-\frac{B_{011}}{2B_{002}}.
\label{zpsicons}
\end{align}
Equating the coefficients of $(\gamma,\Psi,z)$, we obtain the following two equations that determine $\{B_{020},B_{011}\}$. 
\begin{align}
&4 B_{002}B_{020}=B_{011}^2\nonumber \\
&2 B_{002}B_{110}=B_{011}B_{101}\:\:, \label{QS2eq}
\end{align}
exactly the same as the omnigeneity conditions (\ref{omniB1}).  However $y_{002}$ is not required to be zero in omnigeneity. From (\ref{Bsqrseries}) we find that unlike omnigeneity, QS constraint  ($y_{002}=0$), affects the $O(\epsilon)$ term in the expansion of $|\B|$. Thus, the QS constraint starts at a lower order than omnigeneity. 

Note that instead of obtaining three equations from the three coefficients of $(\gamma,\Psi,z)$ in (\ref{zpsicons}), we get only two because the coefficient of $z$ vanishes identically. This feature repeats  in higher orders. 

The next order corrections to the relevant quantities are 
\begin{subequations}
\begin{align}
B^{(3)}&= \sum_{i+j+k=3}B_{ijk}\:\:\gamma^i\Psi^j z^k\\
\Jac&=1+ \ep (B_{100}\gamma) +O(\ep^2)\\
(\dpl B)^{(3)} &= ( 3 B_{003}z^2+B_{201}\gamma^2+2 B_{012}z\Psi  \nonumber\\ 
&+B_{021}\Psi^2+2B_{102}z \gamma +B_{111}\gamma\Psi)\\
(z'_\Psi)^{(1)}&= -\: \frac{B_{011}}{2B_{002}}(-Z_1+Z_2)\:\:,\label{zpsi1}
\end{align}
\end{subequations}
where
\begin{align*}
Z_1=( 3 z^2 B_{003} &+2 z \Psi B_{012}+\Psi^2 B_{021} +2 z \gamma B_{102}\\
+ \gamma \Psi B_{111} &+ \gamma^2 B_{201} )/(B_{011}\Psi +B_{101}\gamma +2B_{002}z) \\
Z_2=( z^2 B_{012} &+2 z \Psi B_{021}+3\Psi^2 B_{030}+  z \gamma B_{111}\\
+ 2\gamma \Psi B_{120} &+ \gamma^2 B_{210} )/(2 B_{020}\Psi +B_{110}\gamma +B_{011}z).
\end{align*}

The QS criteria (\ref{QS0}) is now given by
\begin{align}
2(z'_\Psi)^{(0)}(3 B_{003}z+B_{012}\Psi + B_{102}\gamma)+2B_{002}(z'_\Psi)^{(1)}\nonumber\\
+(2 B_{012}z +2B_{021}\Psi +B_{111}\gamma)=0.
\end{align}
Substituting the expressions for $((z'_\Psi)^{(0)},(z'_\Psi)^{(1)})$ from (\ref{zpsicons},\ref{zpsi1}) we get a quadratic polynomial equation in $(\gamma,\Psi,z)$. Equating the coefficients of $\{\gamma^2,\Psi^2,z^2,z\gamma,z\Psi,\Psi\gamma\}$ in the quadratic polynomial equation to zero we get six equations, of which the four that are linearly independent completely matches the omnigeneity set (\ref{omniB2}). They determine $\{B_{012},B_{030},B_{210},B_{120}\}$.

\subsection{QS with ergodic field lines}
Now let us evaluate and compare the constraints (\ref{BDBpsi}) and (\ref{QH}) with (\ref{QS0}). We will show that (\ref{QS0}) is indeed equivalent to (\ref{BDBpsi}) but is less stringent than (\ref{QH}). We shall start with (\ref{BDBpsi}) and evaluate $B_\Psi$ in the process. Then we shall impose (\ref{QH}) by insisting that the obtained $B_\Psi$ be just a function of $\gamma$. 

Using (\ref{fiz},\ref{fisi}) and (\ref{Bdfi}), we obtain the following expression for the ratio $B_\Psi$
\begin{align}
B_\Psi= \Phi_{,\Psi}-\frac{\Phi_{,z}}{B_{,z}} B_{,\Psi}. \label{hdef}
\end{align}

Since $\partial_z B =O(\epsilon^2)$ and $\partial_\Psi B= O(\epsilon)$, $\ep B_\Psi$ is an $ O(1)$ quantity and 
\begin{align*}
\ep B_\Psi=\frac{2 y_{002}}{(2 B_{002} z+B_{011} \psi +B_{101} \gamma )}+O(\ep).
\end{align*}
The QS constraints can be satisfied iff  \begin{align*}
y_{002}=0 .
\end{align*}
This constraint, equivalent to (\ref{QSreq1}), ensures QS up to first order in the expansion of B. The next order terms in $B_\Psi$ simplifies to 

\begin{align}
B_\Psi=-\frac{2 B_{020}\Psi +B_{110}\gamma +B_{011}z}{B_{011}\Psi +B_{101}\gamma +2B_{002}z} +O(\ep).
\label{Bpsi0}
\end{align}

We note that $(B_\Psi)^{(0)}=(z'_\Psi)^{(0)}$ and therefore demanding it to be independent of $z$ for arbitrary $(\gamma,\Psi)$ requires $(B_\Psi)^{(0)}$ to be a constant to lowest order. This yields the two independent equations for $\{B_{020},B_{110}\}$ derived earlier in omnigeneity (\ref{omniB1}) and QS (\ref{QS2eq}). Here we note that $B_\Psi$ has a denominator which vanishes at the magnetic well minima. The $\{B_{020},B_{110}\}$ constraints make $B_\Psi$ finite. This is true for omnigeneous systems as well. Note also, that by requiring $B_\Psi$ to be independent of $\Psi$ instead of $z$ gives the exact same result. This shows the redundancy in the QS constraint.

Proceeding to the next order, we find that imposing (\ref{BDBpsi}) leads to the exact same set as (\ref{omniB2}). Simplifying the expression of $B_\Psi$ with the help of these equations, we get
\begin{align}
B_\Psi= &-\frac{B_{011}}{2 B_{002}} \label{Bpsiexp} +\\ &+\ep  \left(\frac{B_{011} B_{102}}{2 B_{002}^2} - \frac{B_{111}}{2 B_{002}}-y_{101}- \frac{B_{011} y_{110}}{2 B_{002}}\right) \gamma  \nonumber\\ &+ \ep\left(\frac{3 B_{003} B_{011}^2}{4 B_{002}^3} - \frac{B_{021}}{B_{002}} - y_{011}\right) \Psi.  \nonumber
\end{align}
$B_\Psi$ is manifestly constant along a field line, but is not quasisymmetric according to (\ref{QH}) since the last term in (\ref{Bpsiexp}) is field line label dependent. We can eliminate the $\Psi$ dependence by fixing $y_{011}$. Alternatively, we can fix $B_{021}$,
\begin{align}
4 B_{002}^2 B_{021} = 3 B_{003} B_{011}^2 - 4 B_{002}^3 y_{011}\label{QS5eq}.
\end{align}
 
Therefore, to this order (\ref{QH}) consists of (\ref{QS5eq}) in addition to the set (\ref{omniB2}). This can also be verified through a direct and straightforward implementation of (\ref{QH}). We obtain ten equations by equating to zero the coefficients of $\{z^2,z\:\gamma,z\:\Psi,\Psi\:\gamma,\Psi^2\}$ in the expansions of the two derivatives of $B_\Psi$. We find repetitions and only five out of ten equations are actually independent. These equations determine the third order coefficients of $B$, $\{B_{012},B_{030},B_{210},B_{120},B_{021}\}$. The first four satisfy (\ref{omniB2}) and $B_{021}$ satisfy (\ref{QS5eq}). Note that there are still many free parameters to this order, which can be utilized to enforce QS to higher orders.

\section{A comparison of omnigeneity and QS constraints} \label{sec:comparison}
In our local expansion about a magnetic minima, we have evaluated the omnigeneity constraint (\ref{omnicrit0}) and the QS constraints (\ref{QS0},\ref{BDBpsi}) and (\ref{QH}) systematically up to 3rd order in the distance from the minima. In this section, we compare and summarize the similarities and differences between all the different constraints. 

In section \ref{sec:omniconstraint}, we have shown that the omnigeneity constraint (\ref{omnicrit0}) does not constrain the magnetic field up to first order in $\ep$. To second and third order, it leads to the constraint (\ref{omniB1}) and (\ref{omniB2}) respectively. The constraints manifest as nonlinear algebraic equations for various MHD equilibrium expansion coefficients given in Appendix \ref{appendix B}.

In section \ref{sec:QSconstraint}, we have explored the different forms of QS constraints: (\ref{QS0},\ref{BDBpsi}) and (\ref{QH}), which are all well known in the literature. The last constraint, (\ref{QH}),is the only appropriate constraint when the magnetic field lines are ergodic and is the most general case since ergodic field lines are ubiquitous. The constraints (\ref{QS0},\ref{BDBpsi}) are equivalent but are different from (\ref{QH}) because the latter does not allow field line label dependence of $B_\Psi$. We have shown that (\ref{QH}) is much more stringent than (\ref{QS0}) and adds additional restrictions at each order. 

Compared to omnigeneity, QS constrains the first order magnetic field by requiring $B_{010}=0$, which is not required in omnigeneity. We have shown that (\ref{QH}) imposes additional constraints in order that $B_\Psi$ be independent of $\Psi$. To third order, (\ref{QH}) requires (\ref{QS5eq}) to be satisfied. Omnigeneity is a nonlocal constraint and the constraints involve only field line averaged quantities while QS is a local constraint and involves all three spatial variables. Thus, in a local analysis omnigeneity constraints involve equating coefficients of polynomials involving only $(\gamma,\Psi)$ to zero, while QS involve $z$ as well. Ignoring the redundancies in QS constraints, we see that this nonlocal aspect of omnigeneity leads to fewer omnigeneous constraints compared to QS at any given order.

We found that at each order there are redundancies in the evaluation of the QS constraints. Simply counting the number of equations (based on equating coefficients of various monomials in $(\gamma,\Psi,z)$ to zero) would suggest too many constraints, whereas the actual number of independent equations is significantly smaller. This redundancy seems to be inherent in the structure of the constraints and might also persist in a global analysis.

Besides the above mentioned differences, some of the QS and omnigeneity constraints are quite similar up to third order. We find that the set (\ref{omniB1},\ref{omniB2}) appear in both omnigeneity and QS. The constraints also make $B_\Psi$ finite across the magnetic minima. According to Cary-Shasharina  \cite{caryshasharina1997omnigenity}, a perfectly analytic omnigeneous MHD equilibrium needs to be QS. However, by explicit calculation based on our analytic MHD expansions, we have shown that even analytic omnigeneous equilibrium constraints are still much more relaxed than QS especially (\ref{QH}). We have also shown that in our local non-symmetric MHD equilibrium expansion, we have sufficient independent parameters to satisfy the omnigeneity as well as QS constraints completely to third and possibly higher orders in our local expansion.

\section{Conclusion}
\label{sec:sum}
 In this work, we have explored the QS and omnigeneity constraints analytically. We have shown that at the heart of the omnigeneity constraint lies the requirement that the magnetic well width measured along a field line be independent of the field line label. This is best seen through Abel inversion of the $\Jpl$ integral. This inverse problem approach also clarifies that the magnetic wells do not have to be of the same depth. 
 
 We have then explored the consequences of these constraints on a local 3D MHD equilibrium expansion valid near a magnetic well minima. Instead of assuming an ad-hoc MHD equilibrium we have actually solved for the non-symmetric non-linear MHD equations as derived in Weitzner \cite{weitzner2016expansions}. The expansions can be carried up to arbitrary high order. Based on this expansion, we have evaluated the QS and omnigeneity constraints explicitly. We have shown that omnigeneity is less restrictive than QS. Since we have assumed analyticity in our expansion, the difference in omnigeneity and QS is interesting in its own right in the light of Cary-Shasharina's work \cite{caryshasharina1997omnigenity}. This implies, that within the context of a local expansion, there exists a large class of equilibrium  close to quasisymmetry, which are still omnigeneous (non-analytic as well as analytic) but allow inclusion of quasisymmetry breaking error fields.
  
 Finally, we have shown that a straightforward estimate of the number of equations constraining a QS MHD equilibrium, based on counting the degree of independent monomials in a power series expansion, overestimates the actual number of independent equations to a large extent since there are significant redundancies. Although, we have not presented a global analysis, we expect that the redundancy in the equation system should still persist in a more global setting. In a future work, we shall address this point in details by carrying out the calculation globally near the magnetic axis.

\begin{appendices}
 \begin{center}
      {\bf APPENDIX}
    \end{center}
\renewcommand{\thefigure}{A-\arabic{figure}} 
\numberwithin{equation}{section}
\section{Derivation of Abel Inversions}\label{appendix 0}    
Let us first consider the single well case. Differentiating (\ref{Jplsingle}) with respect to energy we obtain the bounce time as given in (\ref{tausingle}). Next, we divide $\tau(\cE)$ by $ \sqrt{2(B-\cE)} $ and integrate over $\cE$ between $(\Bmin,B)$. Changing the order of integration, we obtain
 \begin{align}
 \int_{B_{\text{min}}}^B \dt\cE\: \frac{\tau(\cE)}{\sqrt{2(B-\cE)}} = \\ 
 \int_{B_{\text{min}}}^B \dt\cB\:\frac{dw}{d\cB}\int_{\cB}^B d\cE \frac{1}{2\sqrt{(B-\cE)(\cE-\cB)}} \nonumber
 \end{align}
Using,
\begin{align}
\int_{b}^B \dt\cE \frac{1}{2\sqrt{(B-\cE)(\cE-\cB)}} = \frac{\pi}{2} - \text{sin}^{-1}{\sqrt{\frac{b-\cB}{B-\cB}}}
\label{intE}
\end{align}
the integral over $\cE$ gives a factor of $(\pi/2)$ and the integral over $\cB$ gives the width of the well $w(B)$. Using the fact that the width vanishes at the minima i.e $w(\Bmin)=0$, we get (\ref{wsingle}).
 The analysis for multiple well is very similar to the single well case. We proceed as before and integrate $\tau(\cE)/\sqrt{2(B-\cE)}$ with respect to $\cE$ from $(\Bmax^{(i)},B)$. Let us now consider the first term in $\tau(\cE)$ from Eq.(\ref{tauE2well}). Changing the order of integration and integrating over $\cE$ first, we obtain
 $$\frac{\pi}{2}\lbr w(B)- \sum_{j=i}^{i+1} w^{(j)}\rbr$$ 
 Using \ref{intE}, the second term in (\ref{tauE2well}) yields
 $$+\frac{\pi}{2}\sum_{j=i}^{i+1} w^{(j)}-\sum_{j=1}^{i+1}\int_{\Bmin^{(j)}}^{\Bmax^{(i)}} d\cB\: \frac{dw}{d\cB}^{(j)}\:  \text{sin}^{-1}{\sqrt{\frac{\Bmax^{(i)}-\cB}{B-\cB}}}$$
 Adding these two and rearranging we obtain (\ref{wdouble}). 

\section{Expression for various coefficients \label{appendix B}}
\subsection*{MHD equilibrium expansion coefficients}
Let us summarize the results of the local MHD equilibrium expansion. We have used Mathematica to carry out the algebra. Through $O(\epsilon)$, we have
\begin{align}
x_{101} = y_{011}, \:\: x_{110} = 2 y_{002}, \:\: 2x_{002} =  p_1+y_{110}
\end{align}  
\begin{align}
\Phi_{002} &= 0, \Phi_{011} = -2 y_{002}, 2\Phi_{020} = -y_{011} \nonumber\\
\Phi_{110} &= -y_{101}, \Phi_{101} = y_{110}.
\end{align}
$\{x_{011},x_{020}\}$ are independent. 
Proceeding to $O(\epsilon^2)$ we obtain,
\begin{align}
2x_{210} = &-2 p_1 y_{002}  + x_{011} y_{011} +y_{011} y_{101} + 2 y_{102} \nonumber \\
- &6 y_{002}\: y_{110} + 2 y_{120} - 4 x_{020}\: y_{200} \nonumber \\
2x_{120} = &x_{011}^2 - 2 p_1 x_{020} + 4 y_{002}^2 + y_{011}^2  \nonumber \\
&+ 2 y_{012}- 6 x_{020}\: y_{110} \\
x_{111} = 
& 6 y_{003} + 2 y_{002} y_{011} + 2 y_{021} \nonumber\\
 &- 2 x_{020}\: y_{101} - x_{011} y_{110}\nonumber\\
  2x_{102} = 
 &2 p_2 + y_{011}^2 + y_{101}^2 - 2 p_1 y_{110} + y_{110}^2\nonumber\\
 &- 12 y_{002} y_{200} +2 y_{210}\nonumber\\
x_{012} = &2 p_1 y_{002} - y_{102}\nonumber\\
2x_{201}= 
 &-6 x_{003} - 2 y_{002} y_{101}+ y_{011} y_{110} + \nonumber\\ 
 &+y_{111} - 2 x_{011} y_{200}\nonumber
 \end{align}
\begin{align}
6\Phi_{003} = &2 p_1^2 - 2 p_2 + 8 y_{002}^2 + y_{011}^2 + 2 y_{012} - 2 x_{011} y_{101}\nonumber\\ 
& - y_{101}^2 + 6 p_1 y_{110} + y_{110}^2 + 12 y_{002} y_{200} - 2 y_{210} \nonumber\\
\Phi_{012} = & p_1 x_{011} - 3 y_{003} + 2 y_{002} y_{011} + x_{011} y_{110} \nonumber\\
\Phi_{102} = & 3 x_{003} + p_1 y_{011} + 2 y_{002} y_{101} + y_{011} y_{110}\\
2\Phi_{021} = & x_{011}^2 + 2 p_1 x_{020} + 4 y_{002}^2 + y_{011}^2 - 2 y_{012} + 2 x_{020}y_{110} \nonumber\\
\Phi_{111}= & 2 p_1 y_{002} + x_{011} y_{011} + y_{011} y_{101} - 2 y_{102} + 2 y_{002} y_{110}\nonumber\\
\Phi_{201} = & y_{011}^2 + y_{101}^2 + y_{110}^2 - 4 y_{002} y_{200} + y_{210} \nonumber\\
3 \Phi_{030} = & 2 x_{011} x_{020} + 2 y_{002} y_{011} + 2 y_{011} y_{020}- y_{021} \nonumber\\
\Phi_{120} = & x_{011} y_{002} +  x_{020} y_{011} +  y_{002} y_{101} - y_{111}/2 \nonumber\\
\Phi_{210}= & 2 y_{002} y_{011} - y_{201} \nonumber
\end{align}
To this order $\{x_{030},x_{003},x_{021}\}$ are independent. 

\subsection*{Coefficients of $B$}

The $B_{ijk}$ coefficients through second order are given by
\begin{align}
B_{100}= &y_{110}, \quad B_{001}= -2 y_{002} \nonumber\\
B_{002}= &(p^2_1 +y_{011}^2- y_{101}^2)/2 -p_2-y_{210}+2 y_{002}^2 +y_{012}\nonumber \\
& -x_{011}y_{101} +2p_1 y_{110}+6 y_{002} y_{200} \nonumber\\
B_{020}= & (p_1+y_{110})x_{020}+y_{002}^2 -y_{012}\\
B_{200}=& (y_{011}^2 +y_{101}^2)/2 +y_{110}^2+y_{210}-4y_{002}y_{200} \nonumber\\
B_{110}=& 2y_{002}(p_1+y_{110}) -2y_{102} \nonumber\\
B_{011}=& x_{011}(p_1+y_{110}) +2y_{002}y_{011} - 6 y_{003}. \nonumber
\end{align}

Some of the third order coefficients are
\begin{align}
    B_{012}=&6 x_{003} x_{011} - 12 y_{004} + 6 y_{003} y_{011} - p_1 y_{102} - y_{102} y_{110}\nonumber\\
    B_{030}= &(p_1+y_{110}) x_{030} + 2 y_{003} y_{011} - 2 y_{022}/3 - 4 x_{020} y_{102}/3  \nonumber\\
    B_{210}= & (p_1 + y_{110}) ((1/2)y_{011} (x_{011} + y_{101})\\
    &+ (y_{102} + y_{120} - 2 x_{020} y_{200})) - 2 (3 x_{003} y_{101} + y_{202})\nonumber \\
    B_{120}= &1/2 (x_{011}^2 + y_{011}^2) (p_1 + y_{110}) -p_1^2 x_{020} - 3 x_{003} y_{011}\nonumber\\
    & + 3 y_{003}  y_{101} + y_{012} y_{110} + p_1 (y_{012} - 5 x_{020} y_{110})\nonumber \\ & - y_{112}
 + x_{020} (2 p_2 + y_{011}^2 + y_{101}^2 - y_{110}^2 + 2 y_{210}) \nonumber\\
    B_{021}=& N_1- N_2(p_1 + y_{110})/(2 y_{101}) \nonumber\\
    N_1= & -((p_1^2 y_{011})/2) - 3 y_{013}- x_{011} y_{102} + 
 y_{011} ( y_{210}\nonumber\\&-x_{020} y_{110}+
 (1/2) y_{011} (2 p_2 - y_{011}^2 + 2 x_{011} y_{101} + y_{101}^2) \nonumber\\
 &+ 
 3 x_{003} (2 x_{020} + y_{110}) + p_1 (3 x_{003} - y_{011} (x_{020} + 2 y_{110})))\nonumber\\
    N_2= & -12 y_{004} + p_1 x_{011} y_{011} - 4 y_{011} y_{021} + 3 p_1 y_{102}  
\nonumber\\    &+ 
  2 x_{011} y_{011} y_{110} + y_{102} y_{110} + x_{011} y_{111}  \nonumber\\
  &+ 2 x_{020} y_{102}-2 (y_{022} + y_{012} y_{200} + y_{202})\nonumber
\end{align}

 \section{Derivations of omnigeneity related identities \label{appendix C}}
\subsection*{Derivation of (\ref{z1diff}) }
We rewrite (\ref{z1s}) using (\ref{deltaz0}) as follows
\begin{align}
&-\sigma |\zbp_0-\zbm_0|\:B_{002}\:\zbs_1 =\nonumber\\
&\quad (B_{003}\:(\zbs_0)^3 +(B_{102}\gamma +B_{012}\Psi)\: (\zbs_0)^2 \label{z1seqn}
\\
&+(B_{201}\gamma^2+ B_{111}\gamma\Psi+B_{021}\Psi^2)\: \zbs_0+ \nonumber\\
&(-\Bb_3+B_{300}\gamma^3+B_{210} \gamma^2 \Psi+ B_{120}\gamma \Psi^2+B_{030}\Psi^3)\nonumber
\end{align}
Summing (\ref{z1seqn}) over $\sigma$ we obtain
\begin{align}
   &-\Delta \zb_0 (\zbp_1-\zbm_1)=\nonumber \\ & B_{003}\left((\zbp_0)^3 + (\zbm_0)^3\right) \nonumber\\
   &+(B_{102}\gamma +B_{012}\Psi)\left((\zbp_0)^2+ (\zbm_0)^2\right) \label{zb1diff}\\ 
   &+(B_{201}\gamma^2+B_{111}\gamma \Psi +B_{021}\Psi^2) \left(\zbp_0+\zbm_0\right)\nonumber \\
   &+2(-\Bb_3+B_{300}\gamma^3+B_{210}\gamma^2 \Psi +B_{120}\gamma \Psi^2+B_{030}\Psi^3).\nonumber
\end{align}
Making use of the following identities
\begin{align*}
   &(\zbp_0)^3+(\zbm_0)^3=\\
   &\quad \quad (\zbp_0+\zbm_0)\left((\zbp_0)^2+(\zbm_0)^2-\zbp_0\zbm_0 \right)\\ 
   &(\zbp_0)^2+(\zbm_0)^2=(\zbp_0+\zbm_0)^2-2 \zbp_0\zbm_0
\end{align*}
to rewrite the first two terms of the right hand side of (\ref{zb1diff}) and collecting terms with $|\Delta \zb_0|^2$ terms we obtain (\ref{z1diff}).
\subsection*{Derivation of (\ref{alf2sum})}
From (\ref{alf2}) we have 
\begin{align}
    &(\alpha^{+}_2-\alpha^{+}_2)\nonumber \\
    &=3 B_{003}\lbr (\zbp_0)^2 -(\zbm_0)^2\rbr+ \\
    &+2B_{002}(\zbp_1-\zbm_1)+2(\Psi\: B_{012}+\gamma\: B_{102})\Delta\zb_0 \nonumber\\
    &=\Delta\zb_0 \lbr 3B_{002}\sum \zbs+2B_{002}\frac{\Delta \zb_1}{\Delta \zb_0} +2(\Psi\: B_{012}+\gamma\: B_{102}) \rbr \nonumber
\end{align}
\end{appendices}
 \begin{acknowledgments}
One of the authors (W.S) acknowledge helpful discussion with Matt Landreman, Josh Burby, Tonatiyuh Vizuet-Sanchez and James Juno. 
This work was supported by the U.S.Department of Energy Grant No. DE-FG02-86ER53223.
\end{acknowledgments}


\bibliography{plasmalit} 
\bibliographystyle{ieeetr}

\end{document}